\documentclass[12pt,twoside]{article}
\usepackage[mathscr]{eucal}
\usepackage{amsmath,amsfonts,amssymb,amsthm}
\usepackage{cite}
\usepackage[matrix,arrow]{xy}
\usepackage{amsthm,amsmath,latexsym,amssymb,amsfonts,amsbsy}
\usepackage{wrapfig}
\usepackage{graphicx}
\usepackage{float}

\usepackage[breaklinks]{hyperref}
\usepackage{amsmath}
\usepackage{amssymb}
\usepackage{braket}
\usepackage[T1]{fontenc}

\usepackage{etoolbox}
\usepackage{multido}

\makeatletter
\def\@@bfil{\leaders \vrule \@height \ht\z@ \@depth \z@ \hfill}
\def\@bLfil{\@@bfil}
\def\@bRfil{\@@bfil}
\def\resetbraceratio{\gdef\@bLfil{\@@bfil}\gdef\@bRfil{\@@bfil}}
\def\setbraceratio#1#2{
  \let\@bLfil\relax
  \multido{\iA=1+1}{#1}{\gappto\@bLfil{\@@bfil}}
  \let\@bRfil\relax
  \multido{\iA=1+1}{#2}{\gappto\@bRfil{\@@bfil}}
}
\def\upbracefill{$\m@th\setbox\z@\hbox{$\braceld$}\bracelu\@bLfil\bracerd\braceld\@bRfil\braceru$}
\def\downbracefill{$\m@th\setbox\z@\hbox{$\braceld$}\braceld\@bLfil\braceru\bracelu\@bRfil\bracerd$}
\makeatother

\usepackage{tikz,pgf}
\usetikzlibrary{shapes}
\usetikzlibrary{calc}
\usetikzlibrary{decorations.pathmorphing}
\usetikzlibrary{decorations.pathreplacing,shapes.misc}
\usetikzlibrary{positioning}
\usetikzlibrary{arrows}
\usetikzlibrary{decorations.markings}

\def\be{\begin{equation}}
\def\ee{\end{equation}}
\def\ba{\begin{array}}
\def\ea{\end{array}}

\def\1{\tilde{1}}
\def\2{\tilde{2}}
\def\3{\tilde{3}}

\newdimen\tableauside\tableauside=.5ex
\newdimen\tableaurule\tableaurule=0.4pt
\newdimen\tableaustep
\def\phantomhrule#1{\hbox{\vbox to0pt{\hrule height\tableaurule
width#1\vss}}}
\def\phantomvrule#1{\vbox{\hbox to0pt{\vrule width\tableaurule
height#1\hss}}}
\def\sqr{\vbox{%
  \phantomhrule\tableaustep

\hbox{\phantomvrule\tableaustep\kern\tableaustep\phantomvrule\tableaustep}%
  \hbox{\vbox{\phantomhrule\tableauside}\kern-\tableaurule}}}
\def\squares#1{\hbox{\count0=#1\noindent\loop\sqr
  \advance\count0 by-1 \ifnum\count0>0\repeat}}
\def\tableau#1{\vcenter{\offinterlineskip
  \tableaustep=\tableauside\advance\tableaustep by-\tableaurule
  \kern\normallineskip\hbox
    {\kern\normallineskip\vbox
      {\gettableau#1 0 }%
     \kern\normallineskip\kern\tableaurule}%
  \kern\normallineskip\kern\tableaurule}}
\def\gettableau#1 {\ifnum#1=0\let\next=\null\else
  \squares{#1}\let\next=\gettableau\fi\next}

\def\be{\begin{eqnarray}}
\def\ee{\end{eqnarray}}
\def\bew{\begin{eqnarray*}}
\def\eew{\end{eqnarray*}}

\def\l[{\phantom.[}

\numberwithin{equation}{section} \makeatletter
\@addtoreset{equation}{section}

\def\be{\begin{equation}}
\def\ee{\end{equation}}
\def\ba{\begin{array}}
\def\ea{\end{array}}

\begin{document}

\begin{flushright}
FIAN-TD-2016-13 \\
\end{flushright}

\vspace{5mm}

\begin{center}

{\Large\textbf{On exact solution  of topological CFT models based  on Kazama-Suzuki  cosets }}

\vspace{5mm}

\vspace{5mm}

{\large Alexander Belavin$^{\;a,c}$ and Vladimir Belavin$^{\;b,c,d}$}

\vspace{0.5cm}

\textit{$^a$ L.D. Landau Institute for Theoretical Physics, \\
 Akademika Semenova av., 1-A, \\ Chernogolovka, 142432  Moscow region, Russia}

\vspace{0.5cm}

\textit{$^{b}$I.E. Tamm Department of Theoretical Physics, \\P.N. Lebedev Physical
Institute,\\ Leninsky av., 53, 119991 Moscow, Russia}

\vspace{0.5cm}

\textit{$^{c}$Department of Quantum Physics, \\ 
Institute for Information Transmission Problems, \\
 Bolshoy Karetny per. 19, 127994 Moscow, Russia}

 \vspace{0.5cm}

\textit{$^{d}$Moscow Institute of Physics and Technology, \\
Dolgoprudnyi, 141700 Moscow region, Russia}

\vspace{0.5cm}

\thispagestyle{empty}


\end{center}
\begin{abstract}
We compute the flat coordinates on the Frobenius manifolds arising on the deformation space of Gepner $\widehat {SU}(3)_k$ chiral rings.
The explicit form of the flat coordinates is important for exact solutions of models of topological CFT 
and 2d Liouville gravity. We describe the case $k=3$,  which is of particular interest
because apart from the relevant chiral fields it contains a marginal one. Whereas marginal perturbations are relevant in different contexts, their analysis requires additional care compared to the relevant perturbations.

\end{abstract}


\section{Introduction}

It was revealed in  \cite{BSZ,BT,Tarn,BDM,VB,BB1,SPO,BYR,Belavin:2014xya} that Frobenius Manifold (FM) structure\cite{Dub} plays an essential role in finding exact solutions
of models of Minimal Liouville gravity (MLG)
\cite{P,KPZ,AlZ,BAlZ}. Apparently this connection  can be also used 
to describe so-called $\mathcal{W}$ extensions of Liouville gravity based on Toda CFTs coupled to $\mathcal{W}_N$ minimal models.
Meanwhile, a solution of Douglas  string equation  \cite{Douglas} required for constructing  the generating function of MLG correlators  has most tractable form in the flat coordinates on the associated FM \cite{VB}. 

An explicit form of the flat coordinates is also necessary for solving models of the topological conformal field theory (TCFT)\cite{DVV}. 
 It was shown in \cite{Gep1} that a certain class of topological conformal field theories, the so-called Gepner models, which obey the symmetry  of  Kazama--Suzuki (KS) cosets
\be\label{coset}
 A(N,k)=\hat{U}(N)_k \times \hat{U}(N-1)_1/ \hat{U}(N-1)_{k+1}\;,
 \ee  
are described in terms of singularity theory, representing a particular class of 
 isolated singularities. 
In this case (see, e.g., \cite{LVW, Mart}) the topological model is described by superpotential  $W_0(x_1,...,x_{n})$, which 
is quasi-homogeneous polynomial  with integer
weights $[x_i]=\rho_i$ and $[W_0]=d$, such that
\be\label{homogen}
W_0(\Lambda^{\rho_i} x_i)=\Lambda^d W_0(x_i)\;. 
\ee 
The main object in TCFT models is correlation functions of chiral fields and their superpartners
$\langle O_{\alpha_1}...O_{\alpha_n}\int\tilde{O}_{\beta_1}...\int \tilde{O}_{\beta_m}\rangle$,  defined completely \cite{DVV} by the {\it perturbed} potential
\be
W(x_i,t^\alpha)=W_0(x_i)+\sum_{\alpha} t^\alpha e_\alpha(x)\;. 
\ee
Here $e_\alpha$ are associated with the perturbation operators and $t_\alpha$ are the coupling constants. 
Possible perturbations are counted by independent elements of the Jacobi ring 
\be
R_0=\mathbb{C}[x_1,...,x_n] /\left\{\frac{\partial W_0[x]}{\partial x_i}\right\}\;.
\ee
The elements $e_\alpha$ represent a particular choice of the basis in $R_0$.
The ring of polynomials $R_0$ is isomorphic to the ring of chiral fields in the corresponding  TCFT models \eqref{coset}. 

The main ingredient of  the  approach \cite{DVV} to the TCFT models  is the {\it deformed} Jacobi ring
\be
R=\mathbb{C}[x_1,...,x_n]/ \left\{\frac{\partial W[x,t]}{\partial x_i}\right\}\;.
\ee
The structure constants of the ring $ R$ is subject of  the following defining relations
\be\label{strconst}
e_\mu e_\nu= C_{\mu \nu}^\lambda e_\lambda \bmod \frac{\partial W[x,t]}{\partial x_i}\;.
\ee
By analogy with the case of one variable $x$ \cite{DVV}, the metric can be written as 
\be\label{metric_g_gen}
g_{\mu\nu}=\underset{x_i=\infty}{\text{Res}} \,\frac{e_\mu e_\nu \Omega}{\prod_i \partial_i W}
\;, 
\ee
where $\Omega(t, x)$ is a so-called {\it primitive form} of Saito \cite{Saito}. 
The important feature of the isolated singularities  \cite{Saito} is that there exits an $\Omega$-form  such that all FM axioms \cite{Dub} are fulfilled
for the deformation parameters space equipped with the metric \eqref{metric_g_gen}.

Lowering $\lambda$ in $C_{\mu\nu}^\lambda$ with the  metric $g_{\mu\nu}$, one obtains 
$C_{\mu\nu\lambda}$ tensor,  whose components are related  to the perturbed three-point functions of the chiral fields  $\langle O_{\mu}O_{\nu}O_{\lambda}\exp{(\sum_{\alpha} s^{\alpha}\int \tilde{O}_{\alpha}})\rangle$.   The exact equivalence  is achieved  by the coordinate transformation  
from the parameters $ t^{\mu}$   to the flat coordinates $ s^{\mu}$. 
The flat coordinate frame, in which the metric is constant, does exist because the Frobenius manifold is flat.                         
As it was shown in  \cite{DVV}, the computation of multi-point correlators is  reduced 
to the computation of the perturbed three-point functions.

An efficient method of computing the flat coordinates for \eqref{coset} models with only relevant perturbations was proposed in
\cite{BGK}. The method is based on the relation between \eqref{coset}  models and Gauss-Manin (GM)
systems of differential equations. This relation allows, in particular, to conjecture the existence of an integral representation constructed from the solutions of  GM systems
for the flat coordinates.
For $A(2,k)$ models this conjecture was checked in \cite{BGK}. In \cite{BB} the generalization 
of the integral representation for the flat coordinates was proposed for models with marginal deformations (defined bellow). 

In this letter we address the problem, discussed earlier in \cite{Saito, Noumi, Blok, Losev,  BGK}, of computing the flat structure $(s^\mu,\Omega)$ on the FMs arising on the deformations of the isolated singularities. We consider an example of $A(3,3)$ model, where the marginal  deformation  appears in the spectrum.  The explicit computation discussed below allows to verify the integral representation for the flat coordinates proposed in \cite{BB}.

\section{Flat coordinates and Primitive form}
\label{sec:genconstr}

For our purposes it is instructive to choose the basis elements $e_\alpha$ to be
 homogeneous 
polynomials 
with weights $\sigma_\alpha$, $[e_\alpha]=\sigma_\alpha$, and
to assign the weights $[t_\alpha]=\epsilon_\alpha$
to the deformation parameters,
according to $\epsilon_\alpha=d-\sigma_\alpha$, so that 
$W(\Lambda^{\rho_i} x_i,\Lambda^{\epsilon_\alpha} t_\alpha)=\Lambda^d W(x_i, t_\alpha)$. 
In our convention $e_1$ is the unity operator, i.e., $\sigma_1=0$. We will use the following standard classification.  
The elements $e_\alpha$ with $\sigma_\alpha<d$, $\sigma_\alpha=d$ and
$\sigma_\alpha>d$ are called relevant, marginal and irrelevant respectively.
Below we discuss the case, where all elements $e_\alpha$ are relevant or marginal.

Since the Frobenius manifold is flat, there exists a coordinate frame $s^\mu$ with 
constant metric $\eta_{\mu\nu}$. 
In order to compute the flat coordinates $s^\mu$ as functions of the perturbation parameters $t=\{t^\alpha\}$ we will use the ideas analogous to those applied  for another class of Ginzburg-Landau models in \cite {Klemm}.

First, we note that if for all elements $\sigma_\alpha\leq 1$, then combining \eqref{strconst} and \eqref{metric_g_gen} one can represent the metric in the form
\be\label{g}
g_{\mu\nu}= C_{\mu \nu}^{\lambda_{max}}h(t) \;,
\ee
where function $h(t)$ can be written schematically\footnote{We say ``schematically'', since strictly speaking the prescription of taking limit in the case of few variables $x_i$ is ambiguous, however we will not use this expression for any explicit computation.} 
\be
h(t)=\underset{x=\infty}{\text{Res}} \,\frac{e_{max}\Omega}{\prod_i \partial_i W}\;,
\ee
and $e_{max}$ is the basis element of maximal weight, $[e_{max}]=\max\{\sigma_\alpha\}$. From the dimensional arguments
one can see that all other elements do not contribute to \eqref{g}. 
Therefore, in order to find the flat coordinates $ s^\mu (t) $ we compute first the structure constants in $t$-frame using \eqref{strconst}, then from the requirement that Riemann tensor $R_{\mu \nu \lambda \sigma}$ for \eqref{g} is zero we derive function~$h(t)$, and finally solving the defining equations 
\be\label{flatdef}
 \nabla_\alpha  \nabla_\beta s^\mu (t)=\Gamma_{\alpha \beta}^\gamma  \nabla_\gamma  s^\mu (t)\;,
\ee
we compute the flat coordinates.
In \eqref{flatdef}  $  \nabla_\alpha $ stands for the covariant derivative
and $\Gamma_{\alpha \beta}^\gamma $ are Christoffel symbols for metric $g_{\mu\nu}$.

Let $M$ be the dimension, so-called Milnor number, of the deformed ring $R$, so that for $t^\alpha$ the index $\alpha=1,...,M$. Using dimensional arguments, one can write the following polynomial expansion of the flat coordinates
in the monomials constructed from the  deformation parameters
\be\label{flat}
s^\mu(t^1,...,t^M)=t^\mu +\sum_{m_\alpha\in \Sigma_\mu} C_\mu(\vec{m}) \prod_{\alpha=1}^M 
\frac{(t^\alpha)^{m_\alpha}}{m_\alpha !}\;,
\ee
where $m_\alpha \in \mathbb{Z}_{\geq 0}$ and  
\be\label{balance}
\Sigma_\mu=\big\{\{m_\alpha\}\big| \sum_{\alpha=1}^M m_\alpha [t^\alpha]=[s^\mu]\big\}\;.
\ee

From the dimensional arguments we note that if marginal operators are present, one can  in principle allow all possible powers of the corresponding coupling constants to appear in \eqref{balance}.
Therefore, we assume that  \eqref{flat} represent infinite series with non-negative powers in marginal deformations.

\section{Flat structure for Gepner models}
\label{Gepner}

The superpotential for the Gepner model with symmetry algebra $A(N,k)$ has the following form
\be
W_0(x_1,...,x_{N-1})=\frac{1}{N+k}\sum_{i=1}^{N-1} q_i^{N+k}\;.
\ee
Here $x_1,...,x_{N-1}$ are elementary symmetric polynomials of $q_1,...,q_{N-1}$.
The chiral ring of the model is isomorphic to the Jacoby ring
\be\label{R0}
R_0(N,k)=\mathbb{C}[x_1,...,x_{N-1}]\big/\{\partial_1 W_0[x]...\partial_{N-1} W_0[x]\}\;.
\ee
One can choose \cite{Gep1} Schur polynomials 
\be\label{schur}
S_{\lambda}[q_1,...,q_{N-1}]=\frac{\det q_i^{N+\lambda_j-j}}{\det q_i^{N-j}}\;,
\ee
as
 a basis in $R^0(N,k)$. The basis elements are enumerated by Young  diagrams  $\lambda$ with $(N-1)$ rows such that $\lambda_1\leq k$.
The dimension of the ring 
\be\label{dim}
\dim R_0(N,k)=\frac{(N+k-1)!}{k!(N-1)!}\;.
\ee
In what follows we are dealing with $A(3,3)$ model
\be\label{W0k3}
W_0(x,y)=\frac{1}{6}( q_1^6+q_2^6 )\;,
\ee
where 
\be
x=q_1+q_2 \quad \text{and} \quad y=q_1 q_2\;.
\ee
The basis elements are 
\begin{align}
e_1\equiv e_{\varnothing}=1\;, \quad 
e_2\equiv e_{\tableau{1}}=q_1+q_2\;, \quad
e_3\equiv e_{\tableau{1 1}}=q_1 q_2\;. \quad
e_4\equiv e_{\tableau{2}}=q_1^2+q_1 q_2+q_2^2\;, \quad \nonumber\\
e_5\equiv e_{\tableau{2 1}}=q_1 q_2(q_1+q_2)\;, \quad
e_6\equiv e_{\tableau{3}}=q_1^3+q_1^2 q_2+q_1 q_2^2+q_2^3\;,\quad
e_7\equiv e_{\tableau{2 2}}=q_1^2q_2^2\;,\quad\nonumber\\
e_8\equiv e_{\tableau{3 1}}=q_1 q_2(q_1^2+q_1 q_2+q_2^2)\;,\quad
e_9\equiv e_{\tableau{3 2}}=q_1^2 q_2^2(q_1+q_2)\;,\quad
e_{10}\equiv e_{\tableau{3 3}}=q_1^3 q_2^3\;.\quad
\end{align}
There is only one marginal deformation $e_{10}$ in this model, all other perturbations $e_{1},...,e_{9}$ are relevant.
The deformed potential  reads
\be\label{R33}
W(q_1,q_2,t)=\frac{q_1^6+q_2^6}{6} +\sum_{l=1}^{10} e_l  t_{l} \;.
\ee

Our first task is to compute structure constants in $t$-frame. To this end we use the definition \eqref{strconst}.
The ideal is generated by $W_x=\partial_{x}W(x,y)$ and $W_y=\partial_{y}W(x,y)$. The main point of the computation is that multiplying $W_{x,y}$ by $x^a$ and $y^b$, where $a,b \in \mathbb{Z}_{>0}$,
we get sufficient number of linear equations
\be\label{syst}
x^ay^bW_x=0\;,\quad  x^cy^dW_y=0\;,
\ee
which can be used to define all necessary multiplication rules $e_i \times e_j$ in the ring $R$.
If the weights of the basis elements are large enough their product contains monomials whose weights exceed $6$. Solving \eqref{syst} for $0\leq a+2b\leq 7$ and $0\leq c+2d\leq 8$, the monomials which do not belong to the basis, considered as unknown variables, can be expressed in terms of the basis elements. In this way we obtain all structure constants.
The results are polynomials in relevant coupling constants and rational functions in the marginal coupling $t_{10}$.

Our next goal is to find explicit form of the metric.
Because of the marginal operator $e_{10}$, the metric is related to the structure constants up to 
some unknown function of $t_{10}$.
It can be written in the form
\be
g_{ij}=h(t_{10}) C_{ij}^{10}\;. 
\ee
We find $h(t_{10})$ from flatness requirement. Calculating the elements of Riemann curvature tensor one finds the following condition
\be\label{f}
2 \left(1-9 x^2\right)^2 h(x) h''(x)-3 \left(1-9 x^2\right)^2 (h'(x))^2-
4 \left(18 x^2+7\right) h^2(x)=0\;,
\ee
where $x=t_{10}$. Using following substitution
\be\label{h}
h(x)=\frac{1}{\left(1-9 x^2\right) y^2(x)}\;,
\ee
one gets
\be
\left(9 x^2-1\right) y''(x)+18 x y'(x)+ 2y(x)=0\;.
\ee
The general solution of this equation is 
\be\label{y}
y(x)= c_1 P_{-\frac{1}{3}}(3 x)+ c_2 Q_{-\frac{1}{3}}(3 x)\;,
\ee
where $P_n(x)$ and $Q_n(x)$ are  Legendre polynomials. Only one solution is regular in $x=0$ and the overall factor is fixed by the initial condition $h(0)=1$. This defines
the integration constants
\be\label{c1c2}
c_1= \frac{\Gamma \left(-\frac{1}{3}\right)^2 \Gamma \left(\frac{2}{3}\right)}{4\ 2^{2/3} \sqrt{3} \pi }\,,\quad
c_2= \frac{\sqrt{3} \Gamma \left(\frac{2}{3}\right) \Gamma \left(\frac{5}{6}\right)}{2 \pi ^{3/2}}\;.
\ee
The lower terms in the series expansion of $h(x)$ are presented below
\be
h(x)=1+7 x^2+\frac{170 x^4}{3}+\frac{21482 x^6}{45}+\frac{184337 x^8}{45}+\frac{14387263 x^{10}}{405}+\frac{20727458824 x^{12}}{66825}+...\;.
\ee

With the explicit metric $g_{\alpha\beta}(t)$ and knowing the structure constants in $t$-frame one can find the solutions of \eqref{flatdef} for the flat coordinates \eqref{flat}
up to arbitrary power of the marginal coupling  $t_{10}$.
Because the expressions of the flat coordinates $t_\alpha$ for small values of $\alpha$ contain many possible monomials in relevant couplings and look cumbersome, we list below  explicit results only for $t_\alpha$ ($\alpha>4$) up to forth order in $t_{10}$
 \begin{align}\label{expl_s10}
s_{10}={}&
t_{10}  \left(1+\frac{7 t_ {10}^2}{3}+\frac{34 t_ {10}^4}{3}\right),\\
s_{9}={}&t_ 9 \left(1+\frac{13 t_ {10}^2}{2}+\frac{1217 t_ {10}^4}{24}\right)\;,\\
s_{8}={}&t_ 9^2 \left(1+5 t_{10}+19 t_ {10}^2+\frac{205 t_ {10}^3}{3}+233 t_ {10}^4
\right)\nonumber\\
&+t_ 8 \left(1+2 t_{10}+4 t_ {10}^2+
\frac{34
t_ {10}^3}{3}+28 t_ {10}^4\right)\;,\\
s_{7}={}&t_ 9^2 \left(6 t_{10}+84 t_ {10}^3\right)+t_ 7 
\left(1+5 t_ {10}^2+\frac{106 t_ {10}^4}{3}\right)\;,\\
s_{6}={}&t_ 6 \left(1+\frac{3 t_ {10}^2}{2}+\frac{79 t_ {10}^4}{8}\right)+t_ 5 \left(t_{10}+\frac{29 t_ {10}^3}{6}\right)+t_ 8 t_ 9 \left(1+5 t_{10}+\frac{33 t_ {10}^2}{2}+\frac{355 t_ {10}^3}{6}+\frac{1499 t_ {10}^4}{8}
\right)
\nonumber\\
&+ t_ 7 t_ 9 \left(1+\frac{33 t_ {10}^2}{2}+\frac{1499 t_ {10}^4}{8}
\right)+
t_ 9^3 \left(\frac{5}{6}+\frac{35 t_{10}}{3}+\frac{125 t_ {10}^2}{4}
+\frac{4795 t_ {10}^3}{18}+\frac{26675 t_ {10}^4}{48}\right)
\;, \\
s_{5}={}&t_ 5 \left(1+3 t_ {10}^2+19 t_ 
{10}^4\right)+t_ 6 \left(2 t_{10}+\frac{26 t_ {10}^3}{3}\right)+t_ 8
t_ 9 \left(2+8 t_{10}+30 t_ {10}^2+\frac{296 t_ {10}^3}{3}+334 t_ 
{10}^4\right)\nonumber\\
&+t_ 7 t_ 9 \left(8 
t_{10}+\frac{296 t_ {10}^3}{3}\right)+t_ 9^3 \left(\frac{8}{3}+\frac{32 t_{10}}{3}+104 t_ 
{10}^2+\frac{2144 t_ {10}^3}{9}+\frac{5624
t_ {10}^4}{3}\right)\;,\\
s_4={}&t_4+ t_6 t_9+ t_5 t_9+\frac{1}{2} t_8^2+ t_7 t_8+2 t_7 t_9^2+4 t_8 t_9^2+\frac{7}{2} t_9^4\;,\\
s_3={}&t_3+2 t_6 t_9+3  t_8 t_9^2+3 t_7 t_9^2+t_8^2+ 2 t_9^4\;, \\
s_2={}&t_2+\frac{63 t_9^5}{8}+\frac{7}{2} t_7 t_9^3+\frac{21}{2} t_8 t_9^3+3 t_5 t_9^2+\frac{3}{2} t_6 t_9^2+\frac{3}{2} t_7^2 t_9+\nonumber\\
&+3 t_8^2 t_9+t_4 t_9+3 t_7 t_8 t_9+t_6 t_7+t_5 t_8+t_6 t_8\;,\\
s_1={}& t_1+\frac{119 t_9^6}{15}+\frac{28}{3} t_7 t_9^4+14 t_8 t_9^4+\frac{7}{3} t_5 t_9^3+\frac{13}{3} t_6 t_9^3+\frac{13}{2} t_8^2 t_9^2+t_3 t_9^2+t_4 t_9^2+\nonumber\\& 
+2 t_5 t_7 t_9+7 t_7 t_8 t_9^2+
2 t_5 t_8 t_9+2 t_6 t_8 t_9+\frac{t_7^3}{3}+\frac{t_8^3}{3}+t_7 t_8^2+t_5 t_6+t_4 t_8-\frac{t_6^2}{2}\;.
\label{expl_s1}
\end{align}

\section{Comparing to the alternative computation}
In \cite{BB} an alternative approach was used for the computation of the flat coordinates and the primitive form.  In the case $k=3$ the primitive form looks like 
\be
\Omega=\lambda(t_{10})dx dy\;.
\ee 
It was obtained that 
\be
\lambda(t_{10})= {}_2F_{1}^{-1}\left(\frac{1}{6},\frac{1}{3};\frac{1}{2}\bigg|9t_{10}^2\right)\;.
\ee
This alternative approach is based on an assumption about the integral representation 
for the flat coordinates.
To compare the results of the two approaches and to verify the conjecture of \cite{BB} we can use the relation
\be\label{h-s}
h(t)=\frac{\partial s^{10}(t)}{\partial t^{10}}\;.
\ee
To justify  this relation we use the coordinate transformation for  the metric 
\be
{g}_{\mu\nu}=h(t) C_{\mu\nu}^{10}=\eta_{\alpha\beta}\frac{\partial s^\alpha(t)}{\partial t^{\mu}}\frac{\partial s^\beta(t)}{\partial t^{\nu}}\;.
\ee
Using this relation for $\mu=1$ and  $\nu=10$  and taking into account that $C_{1,10}^{10}=1$, we obtain 
\be\label{h-eta} 
h(t) =\eta_{\alpha\beta}\frac{\partial s^\alpha(t)}{\partial t^{1}}\frac{\partial s^\beta(t)}{\partial t^{10}}\;.
\ee
On the other hand,  
\be
\frac{\partial s^{\alpha}(t)}{\partial t^{1}}=\delta_{\alpha 1 }\;.
\ee
Indeed, both $t^1$ and $s^1$ correspond to the unity element of the ring $R$. Inserting this relation into \eqref{h-eta} one finds \eqref{h-s}.
Taking the explicit expression for $ s^{10}$ from \cite{BB} we arrive to the expression
\be
h(t_{10})=\frac{{}_2F_{1}\left(\frac{2}{3},\frac{5}{6};\frac{3}{2}\bigg|9t_{10}^2\right)}{{}_2F_{1}\left(\frac{1}{6},\frac{1}{3};\frac{1}{2}\bigg|9t_{10}^2\right)}+t_{10}\frac{\partial }{\partial t^{10}}\frac{{}_2F_{1}\left(\frac{2}{3},\frac{5}{6};\frac{3}{2}\bigg|9t_{10}^2\right)}{{}_2F_{1}\left(\frac{1}{6},\frac{1}{3};\frac{1}{2}\bigg|9t_{10}^2\right)}\;,
\ee
which coincides with \eqref{h}.

The expressions for the flat coordinates \eqref{expl_s10}-\eqref{expl_s1} coincide with the results obtained in \cite{BB} and hence confirm the integral representation for the flat coordinates proposed there. 

\vspace{10mm} 

\noindent \textbf{Acknowledgements.} We thank  D.~Gepner and L.~Spodyneiko 
for useful discussions.  The work was performed with the financial support of the Russian Science Foundation (Grant No.14-12-01383).


%

\providecommand{\href}[2]{#2}\begingroup\raggedright

\endgroup

\end{document}